\documentclass[journal,10pt,letterpaper,final,twocolumn]{IEEEtran}%
\usepackage{amsmath}
\usepackage{amsfonts}
\usepackage{cite}
\usepackage{amssymb}
\usepackage{graphicx}
\usepackage[usenames,dvipsnames]{color}
\usepackage{epstopdf}
\usepackage[all]{xy}
\usepackage[]{mcode}

\usepackage{diagbox}

\providecommand{\U}[1]{\protect\rule{.1in}{.1in}}
\pdfoutput=1
\providecommand{\U}[1]{\protect\rule{.1in}{.1in}}

\newcommand{\qed}{\nobreak \ifvmode \relax \else
      \ifdim\lastskip<1.5em \hskip-\lastskip
      \hskip1.5em plus0em minus0.5em \fi \nobreak
      \vrule height0.75em width0.5em depth0.25em\fi}

\begin{document}

\title{Smart Buildings Enabled by 6G Communications}
\author{Shuping Dang, \textit{Member, IEEE}, Guoqing Ma, \textit{Student Member, IEEE}, Basem Shihada, \textit{Senior Member, IEEE}, Mohamed-Slim Alouini, \textit{Fellow, IEEE}
  \thanks{The authors are with Computer, Electrical and Mathematical Sciences and Engineering (CEMSE) Division, King Abdullah University of Science and Technology (KAUST), 
Thuwal 23955-6900, Kingdom of Saudi Arabia (e-mail: \{shuping.dang, guoqing.ma, basem.shihada, slim.alouini\}@kaust.edu.sa).}}
\maketitle

\begin{abstract}
Smart building (SB), a promising solution to fast-paced and continuous urbanization around the world, includes the integration of a wide range of systems and services and involves the construction of multiple layers.  SB is capable of sensing, acquiring, and processing a very large amount of data as well as performing appropriate actions and adaptation. Rapid increases in the number of connected nodes and thereby the data transmission demand of SB have led to conventional transmission and processing techniques becoming insufficient to provide satisfactory services. In order to enhance the intelligence of SBs and achieve efficient monitoring and control, sixth generation (6G) communication technologies, particularly indoor visible light communications (VLC) and machine learning (ML), are required to be incorporated in SBs. Herein, we envision a novel SB framework featuring a reliable data transmission network, powerful data processing, and reasoning abilities, all of which are enabled by 6G communications. Primary simulation results support the promising visions of the proposed SB framework.
\end{abstract}

\section*{Introduction}
Urbanization has been sharply accelerated in recent decades, and the United Nations Population Fund (UNFPA) has forecast that around 60\% of the global population will live in urban areas by 2030 \cite{6525602}. Feasible solutions to settle such a large number of people are being sought in order to provide sustainable and high-quality standards of life and efficient resource management in urban areas. Among a number of potential solutions, smart building (SB) has many advantages. SB is a high-profile concept belonging to the category of smart cities, and has attracted researchers' attention with  advances in artificial intelligence (AI) and the Internet of Things (IoT) \cite{7883984}. It integrates a wide range of systems and services into a unified platform. SBs are able to perceive the environment, acquire, and process relevant data, as well as respond to changes of the environment and/or users' needs with a high degree of intelligence and autonomy \cite{8291121}. The aforementioned abilities allow SB to provide various intelligent indoor services for residents (e.g., tracking, navigating, positioning, and downloading). Moreover, SB can also monitor and control the global operating status.

To achieve such complex functionality, the framework of an SB must be constructed over a multi-layer structure consisting of the sensing layer, network layer, semantic layer, software layer, processing layer,  reasoning layer, and service layer. Note that herein we intend not to include an interactive interface for user interaction in the multi-layer structure, since this is an independent functional module and can, to some extent, be regarded as a part of the external environment. The multi-layer structure of an SB with an interactive interface is illustrated in Fig. \ref{smartbuildingstructure}. In order to fully exploit the advantages of the SB and provide satisfactory services to its residents, reliable connectivity and efficient information processing infrastructure for data transmission and distributed processing are indispensable. Consequently, high reliability and intelligence are  crucial technical barriers hindering the practical use of SBs  \cite{6525602}.

To overcome these barriers, we here propose a novel framework for  SBs enabled by sixth generation (6G) communication technologies, harnessing indoor visible light communications (VLC) and machine learning (ML) \cite{dang2020should}. An indoor VLC module is implemented for reliable and massive data transmission in order to ensure that the raw data collected by distributed sensors is received and used effectively throughout the entire SB framework. As a by-product, indoor VLC can also satisfy communication demands as a supplementary of radio frequency communications (RFC) from residents living in the SB. ML is mainly employed to enhance the intelligence of the SB and enable real-time smart control. The framework presented herein is validated by simulation results and found to be a feasible solution by which the two main barriers currently handicapping the development of SBs may be overcome.

\section*{Motivations}

\begin{figure}[!t]
\centering
\includegraphics[width=2.5in]{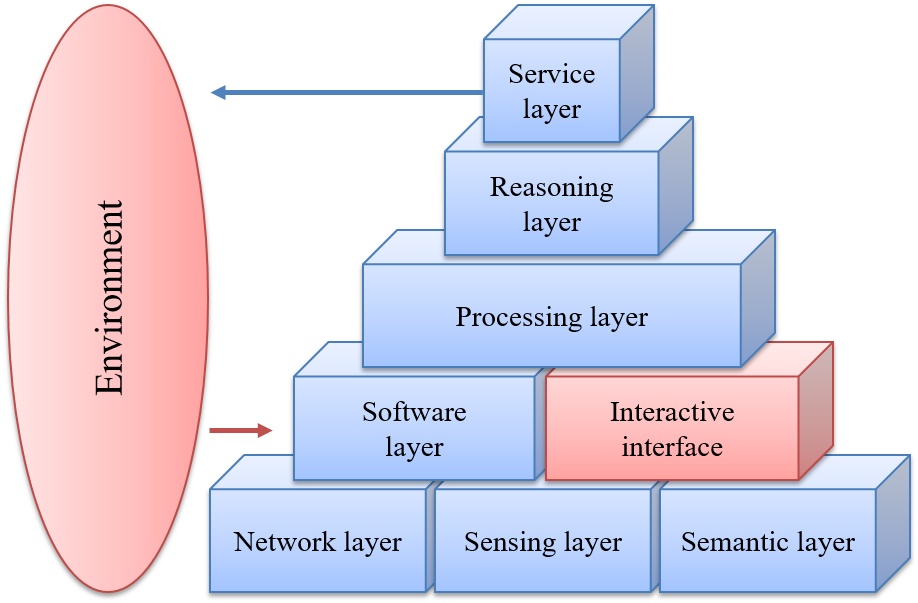}
\caption{Multi-layer structure of the smart building with its interactive interface.}
\label{smartbuildingstructure}
\end{figure}

In order to provide SBs with high reliability and intelligence, we aimed to equip SBs with two key technologies in 6G communications: indoor VLC and ML, and propose a complete framework with details of all key fabrics. As shown in Fig. \ref{smartbuildingstructure}, the framework of the SB can be split into seven functional layers and an interactive interface, in which the sensing and network layers are partially supported by indoor VLC in combination with other communication approaches and the semantic, software, processing, and reasoning layers are strongly associated with ML techniques. The service layer and interactive interface are also directly affected by ML techniques, which conduct and display ultimate outcome outputs using a complex reasoning procedure based on ML algorithms. Using such a framework, the SB, ML, and indoor VLC are intricately interconnected and form a dynamic and holistic system. The benefits and motivations of the proposed framework are detailed as follows. 

Since the most important feature of VLC is the availability of large and unregulated bandwidth, indoor VLC is a promising approach to handle the massive data transmission relevant to SBs in the 6G era, where there exist a huge number of sensors for data collection \cite{retamal20154}. Because the security of SBs takes priority over other design metrics, indoor VLC is able to offer secure and reliable connections against jamming, eavesdropping and other cyber attacks via the construction of a physically isolated network. Aside from reasons of security, the reduction of energy consumption is also a key metric for SB, and, since all data transmissions are piggybacked into illumination, indoor VLC can help reduce the energy consumption required for data transmission.  As a by-product, indoor VLC can also help to offload the cellular and household communication demands of residents and improve the quality of service (QoS) when coexisting with RFC. Table \ref{comparison} provides a  comprehensive qualitative comparison between RFC and VLC.

\begin{table}[!t]
\renewcommand{\arraystretch}{1.3}
\caption{Comparisons between RFC and VLC. Part of the information is extracted and summarized from \cite{7239528}}
\label{comparison}
\centering
\begin{tabular}{c|c|c}
\textbf{Key Indices} & \textbf{RFC} & \textbf{VLC}\\
\hline\hline
Spectrum & $<300$ GHz & $428\sim 750$ THz\\
\hline
Regulation & Licensed & Unlicensed\\
\hline
Energy efficiency & Low & High \\
\hline
Spatial reuse rate & Medium & High \\
\hline
Security & Medium & High\\
\hline
Coverage & Large & Small\\
\hline
Mobility & High & Low\\
\hline
Complexity & High & Low\\
\hline
Implementation & Medium & Easy\\
\hline
Maintenance & Medium & Easy\\
\hline
Design flexibility & High & Low\\
\hline
Signal type & Bipolar & Unipolar\\
\hline
Multi-path fading & Severe & Trivial\\
\hline
Shadowing & Medium & Severe\\
\hline
Noise/Interference & Low & Severe\\
\hline
EMI level & High & Trivial \\
\hline
\end{tabular}
\end{table}

Additionally, in the context of 6G, high intelligence is a key feature of SBs. It indicates that an adaptive mechanism needs to be implemented such that SBs can learn from collected data and improve over time with a high degree of autonomy. Due to real-time control requirements and the vast volume of data collected in SBs, traditional processing techniques are no longer competent, and ML stands out as having uniquely advantageous  capabilities to deal with big data in SBs \cite{8291120}. ML is also computationally efficient and, thus, suitable for volatile environments, from which it is able to extract useful information from massive observed data to make decisions and improve setting parameters in an iterative manner. Moreover, ML is able to conduct pattern recognition and prediction, as well as resource allocation by utilizing historical data, which are necessary for extracting contextual information and providing proactive actions when considering long-term objectives. In the semantic layer, ML can interpret users' demands and allow demand inputs via voice or other customized interactive approaches by pattern recognition. ML is expected to be used throughout the software, processing and reasoning layers as a kernel from which to construct a complete adaptive mechanism and thus improve the services provided by SBs according to established objectives. In the service layer, ML supports a large number of auxiliary functions (e.g., energy saving, space planning, resource coordinating, indoor navigating, positioning, and smart alerting).

\begin{figure}[!t]
\centering
\includegraphics[width=3.5in]{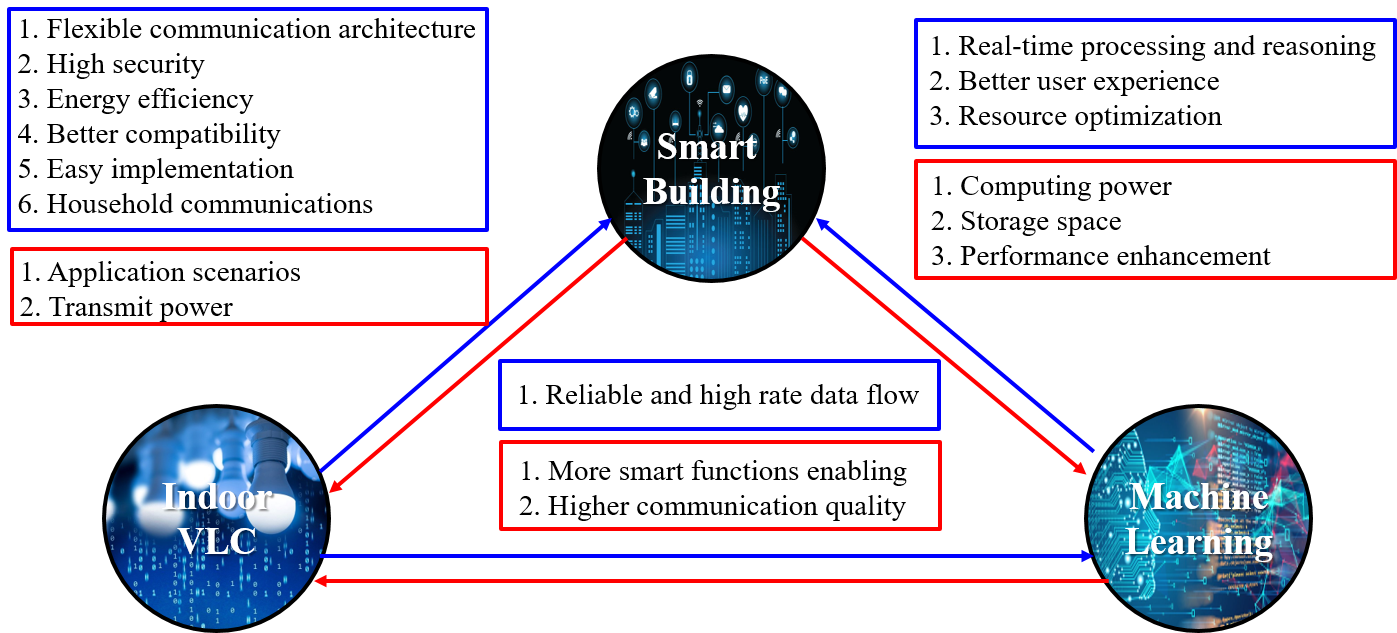}
\caption{Interdependent relationships among the smart building, indoor visible light communications, and machine learning in the 6G era.}
\label{interdependent}
\end{figure}

Although indoor VLC and ML enable SBs in the 6G era, SBs also reciprocally enable the success of indoor VLC and ML. The physical properties of VLC are suited to indoor scenarios, and SBs provide such an application scenario. Furthermore, SBs provide reliable and sufficient transmission power for VLC. In the case of ML, SBs offer sufficient power for computing and provide a large volume of storage space and processors to carry out rapid big data analytics. We depict the interdependent relation among the SB, indoor VLC, and ML in Fig. \ref{interdependent}.

\section*{Proposed Framework}
In this section, we present details of the multi-layer framework for SBs in the context of 6G together with its interactive interface. The functional layers of the proposed framework integrate state-of-the-art sensing, communications, networking, and processing techniques.

\subsection*{Interactive Interface}
The interactive interface is designed to enable interactions between human users and intelligent systems embedded within SBs. The interactive interface should be designed in a human-centric manner. Accordingly, two kinds of interactive interfaces, the fixed control panel and mobile control terminal, may be provided in SBs depending on users' accessibility and usage habits. The former is installed by upgrading pre-existing smart electricity meters or centralized temperature controllers. The latter can be downloaded online as an app to smartphones and/or tablets. Using the interactive interface, residents can monitor the security status of spaces of interest and obtain resource usage profiles, as well as other basic information. As a bidirectional system, users can also provide feedback, submit requests for services, and report issues for attention. The interactive interface is directly linked to the semantic layer.

\subsection*{Semantic Layer}
In the semantic layer, original user inputs are treated as raw data and mapped to machine languages. Using ML in the semantic layer, voice and gesture recognition enable the extraction of contextual information and interpret users' demands accurately. The extracted information and interpreted demands from users are pre-processed and compressed before being sent to the network layer for transmission. Another important function of the semantic layer is to label user feedback such that  emergency feedback can be transmitted and processed with priority over other non-emergency messages.

\subsection*{Sensing Layer}
With the exception of a small portion of information sent from the interactive interfaces by users, most data throughput originates from the sensing layer. Signals generated in the sensing layer contain a variety of observable environmental data, including security, safety, temperature, humidity, space occupancy, electricity usage, water usage, other optical, and acoustic information. To collect such a variety of environmental information and ensure an accurate understanding of the surrounding environment, a large number of sensors are indispensable. In order to reduce the implementation cost of the proposed framework, one should try to reuse existing facilities and instruments and only require the installation of new sensing modules and devices if necessary.

\subsection*{Network Layer}
The network layer supports the transmission and reception among the functional layers. Additionally, because cloud technology and other distributed computing architectures are adopted in the following layers, the network layer requires the construction of a secure and reliable connection among a large number of distributed controllers and processors \cite{mumtaz2014smart}. Meanwhile, the accessibility to the Internet and cellular networks is a basic demand and should also be supported in the network layer.  Indoor VLC, owing to its advantageous properties, has been employed as the protagonist in the network layer. However, to overcome some of the drawbacks of indoor VLC and optimize the communication service provided, two further supporting roles are required: RFC and power line communications (PLC) \cite{7239528}. RFC can be employed for mobile data transmission and provides a supplementary transmission mechanism via mode selection. Meanwhile, due to its low cost of deployment, PLC relying on the existing power supply infrastructure in SBs is an attractive approach to connect light emitting diode (LED) transmitters and is adopted in the proposed framework as a networking backbone.

\subsection*{Software Layer}
The software layer is employed as an interface to receive raw data from the network layer and provide software platforms to process and store these data. In particular, the software layer should support  interactions with the external environment and the service layer by defining I/O interfaces and activating control programs. In order to achieve these functions, first, a powerful database must be constructed and used to store historical data from various sensors and interactive interfaces. Additionally, cloud and distributed computing should be supported, since the hardware architecture adopted in the proposed framework is based on distributed controllers and processors.

\subsection*{Processing Layer}
The processing layer is utilized to pre-process large amounts of raw data, which are presented in different formats and structures, thereby minimizing data redundancy and restoring missing data where possible. Dimensionality reduction is another important function of the processing layer, by which the system aims to maintain the validity of sensory information using a minimum number of variables by means of data redundancy elimination. To achieve this goal, the processing layer must be able to extract the features of different data and perform appropriate selection and projection. ML techniques can also play a role in this functionality. In short, processed data must be ready in unified and appropriate forms for use by ML techniques in the reasoning layer.

\subsection*{Reasoning Layer}
The reasoning layer is the intelligence kernel in the proposed framework, supporting all intelligent functions and services in the SB. In this layer, ML is the absolute protagonist and performs diverse intelligent reasoning based on various application requests. In essence, ML in the reasoning layer provides an adaptive mechanism capable of learning from historical data when the learning objectives are specified by the users or system designers. All intelligent functions and services, as well as the status of the entire SB, are controllable by a set of parameters that  can be changed in the reasoning layer according to input data containing  demands and sensory information from the users. Between the input data and output parameter set, appropriately designed ML algorithms suitable for different scenarios can adaptively optimize its parameters according to output feedback. After having been trained by several training datasets, the reasoning layer will be capable of producing optimized output parameters for the service layer.

\subsection*{Service Layer}
The service layer consists of actuators controlled by the output parameters from the reasoning layer and can therefore change the status of the SB. For a typical SB, these actuators include, but are not limited to, temperature and humidity controllers in air-conditioning systems, smart switches of a variety of electric apparatuses, dimming controllers, smart stereos, safety alarms, video surveillance cameras, and LED transmitters for VLC. Using this smart framework and the smart functions supported by ML, all services in SBs are expected to be continuously improved in the long term by iterative training using new datasets.

\subsection*{Promising Visions}

When indoor VLC and ML are integrated by the proposed framework detailed above,  new features emerge. By harnessing these features, more advanced services can be provided for users, and the operational efficiency of  SBs can be significantly improved in the 6G era. First, by employing VLC in combination with other communication approaches (e.g., RFC and PLC integrating optical sensing), environmental parameters in SBs can be accurately detected and transmitted to higher layers. ML supports the rapid processing of such large amounts of data and enables the display of real-time monitoring information to users on interactive interfaces. In this way, an accurate profile of the indoor environment can be constructed in real time. When providing accurate information regarding the indoor environment in real time,  by-products include various location-aware services, including localization and navigation. 

\begin{figure*}[!t]
\centering
\includegraphics[width=7.0in]{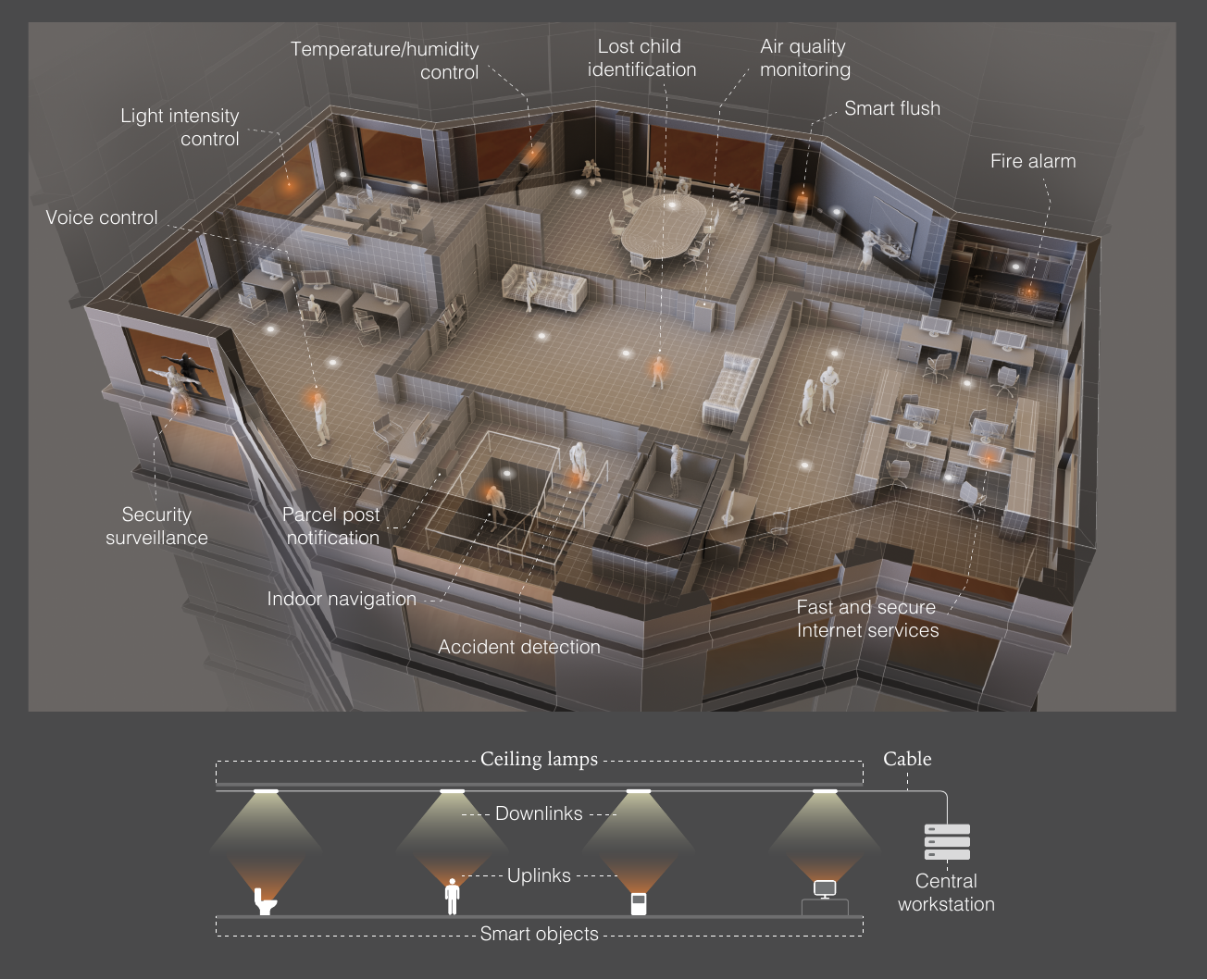}
\caption{A pictorial illustration of the anticipated application scenarios of SBs in the 6G era jointly supported by VLC and ML (Illustration created by Ivan Gromicho. Scientific Illustrator, Research Communication and Publication Services. Office of the Vice President for Research. King Abdullah University of Science and Technology).}
\label{picillu}
\end{figure*}

Despite excellent observability, the joint application of indoor VLC and ML in SBs also results in far better controllability, benefiting from the high-rate transmission and powerful reasoning capability provided by both techniques. Consequently, the indoor environment in SBs can be adjusted to be more comfortable for residents in a smart and rapid manner. A pictorial illustration of the anticipated application scenarios of SBs coupled by 6G communication techniques is presented in Fig. \ref{picillu}.

\section*{Case Study and Validation}
To ensure rigor, we use a simple indoor localization example to evaluate the performance of combined VLC and ML algorithms. This involved setting a simulation platform in a cuboid room with width, length, and height of 10 m, 10 m, and 3 m, respectively. To simulate the scenario incorporating both VLC and ML, we further assumed the presence of commercial LEDs, which were modeled by point light sources installed on the ceiling separated from each other by 1 m. This configuration is similar to that shown in \cite{7408994}, in which four white spotlight LEDs were installed in a cuboid room. Consequently, 81 LEDs were installed on the ceiling. Moreover,  four WiFi access points (APs) were assumed to be installed on the ceiling at a distance of 2.5 m to the walls and separated by 5 m, which is even more than a usual configuration in practice. In order to achieve a comprehensive comparison between the performance of WiFi and VLC, we further assumed another scenario in which  four LEDs were installed in the same manner as the WiFi APs. We denote the results for this configuration as VLC-4 and  the results when utilizing 81 LEDs as VLC-81.

We utilized received signal strength (RSS) based algorithms to predict the locations of receivers. Normally, these algorithms require the pre-installment of receivers to collect RSS data and build datasets. Thereafter, RSS based algorithms can manipulate the dataset in order to predict the location of a new receiver with a new RSS. Without losing generality, the pre-installed receivers are assumed to be located at a height of 1 m above the floor, which is typically the height of a phone held by a human. We set a grid of 99 by 99 receivers in which each was separated from adjacent receivers by 0.1 m with a received field of view (ROV) of 0.7854, and the azimuth angle of each receiver was randomly chosen from -60\textdegree$~$to 60\textdegree. The RSS dataset for VLC was generated by CandLES, a communication and lighting emulation platform   \cite{candles}.

To generate datasets for training purposes, we ran the simulation for fifty times for both the VLC scenario and WiFi scenario. Instead of training a single ML model in a two-dimensional space, we established two ML models to enable separate localization on the x-axis and y-axis. Accordingly, for each ML model, we used 50 by 99 instances for training. After training the models for VLC and WiFi, we applied the same settings to generate new datasets for testing the trained models. Specifically, we utilized the accuracy rate corresponding to different prediction error distances (PEDs) as a measure to evaluate different localization approaches. The PED is defined as the Euclidean distance between the predicted location and the authentic benchmark, and the accuracy rate is consequently defined as the number of predicted locations whose distances to the authentic benchmarks are smaller than the PED. A larger accuracy rate corresponding to a smaller PED thus yields a more accurate localization system. Since we generated the training datasets for each receiver with a 0.1 m separation between adjacent receivers, the precision of the simulated system was 0.1 m.

\begin{figure}[!t]
\centering
\includegraphics[width=3.5in]{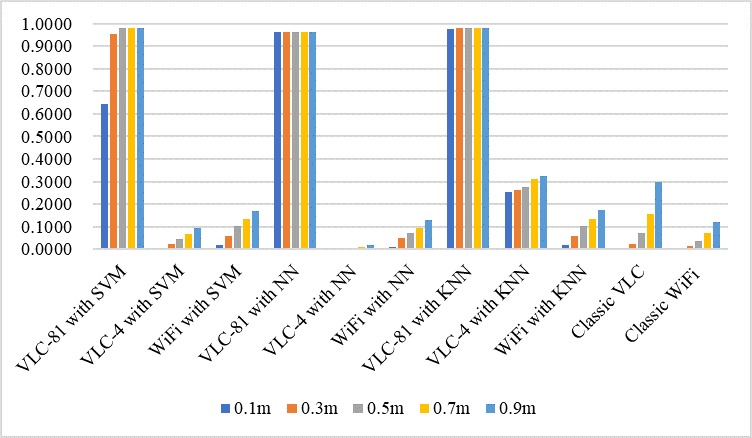}
\caption{Accuracy rates for different prediction error distances (PEDs) under various conditions.}
\label{AccuracyRate}
\end{figure}

The accuracy rates for different PEDs using various localization methods are presented in Fig. \ref{AccuracyRate}. To be comprehensive, we employed three representative ML algorithms: support vector machine (SVM), neural network (NN), and K-nearest neighbors (KNN) to assist  localization data processing.

As shown in Fig. \ref{AccuracyRate}, different ML algorithms result in different accuracy gains in the localization systems, and the superiority among different ML algorithms could change in terms of the required PED and affordable system complexity. It is also evident that accuracy rates increase with increasing PED for all adopted ML algorithms, which is consistent with our expectation. As shown by the results, the performance of VLC-4 with KNN is better than that of WiFi with KNN. However, the performance of VLC-4 is worse than that of WiFi, when applying the SVM and NN algorithms. The reason is that VLC has a better directionality than WiFi, which indicates that the received signal at different locations of VLC has a higher degree of independence. Additionally, VLC-81  generates the best results when equipped with all three algorithms, and the results generated by the SVM and KNN algorithms converge when PED becomes large.

Considering the physical size of a human, 0.3 m is deemed an applicable value of PED for practical indoor localization systems. The corresponding accuracy rates produced by VLC based localization assisted by multiple ML algorithms are greater than 95\%, justifying the feasibility and promising future directions for such a joint approach in SBs.

\section*{Challenges and Potential Future Research Directions}
As a prototype framework, much can still be done to further promote and improve the framework in practice; this should form the basis of future work for 6G communications. In this regard, we articulate several challenges and potential future research directions.

\subsection*{HetNet and Interference Management}
A heterogeneous network (HetNet) architecture should be adopted in the network layer of the proposed framework, consisting of VLC, RFC, and PLC, which inherently increases the efficiency of the entire network layer. However, coordination among heterogeneous communications is not a trivial task \cite{7402263}.  Gateway design and compatibility should be given special attention, and relevant standardization works should also be considered in order to support the HetNet architecture of SBs in practice.

From Table \ref{comparison}, we know that interference has a more severe  impact on VLC than RFC. Therefore, to ensure the performance superiority of the indoor VLC in the proposed framework, interference mitigation technologies must be applied to maintain interference levels below a certain threshold. Since LED transmitters are the main source of interference in the indoor environment, appropriate VLC network deployment and LED transmitter placement are crucial for alleviating interference. Moreover, resource allocation and multiple-input and multiple-output (MIMO) beamforming could also be promising approaches for mitigating interference and optimizing the overall performance \cite{8515272,8119947}. These are still being researched for SBs.

\subsection*{Architecture of ML Algorithms}
One should note that ML is a generic concept representing a package of different learning techniques, which can together be classified into  major domains, such as supervised learning, unsupervised learning, semi-supervised learning, and reinforcement learning according to whether or not the training datasets are labeled. ML can also be divided according to training method used, for example, SVM, NN, KNN, decision tree (DT), and logistic regression (LR). Current consensus holds that there is no learning technique able to assess all cases, as all such techniques have different pros, cons, and applicable scenarios \cite{8755300}. For example, the KNN algorithm is a non-parametric ML algorithm, which implies that it does not require a model training process before making inferences on new data. However, as the dataset becomes larger over time, the inference complexity for this type of non-parametric ML algorithm will increase sharply. On the contrary, for algorithms such as NN and SVM, a model training process is required, but their inference complexity maintains fixed. Therefore, real-time applications can be satisfactorily fulfilled by the meticulous design of a parametric algorithm. In spite of the advantageously short computing time, parametric algorithms designed based on historical records might not produce accurate inference on new instances  because user preferences may vary over time, causing the distribution of collected data to change over time; this is termed  `concept drift'.

Finally, the architecture designs of conventional algorithms and deep learning networks are also of great importance in terms of the functionality and performance of the proposed framework, including how many parameters should be trained and which types of structures should be placed in the training network. Further research pertaining to the selection of ML techniques and deep leaning neural network architecture should be carried out.

\subsection*{SB-Edge-Cloud Computing Architecture}
In most cases, data are collected locally and the corresponding ML models for processing these data are also trained locally. However, as the amount of data available grows rapidly, local computing power could become insufficient to cope with the increasing complexity of ML models. For this reason, SB-Edge-Cloud computing architecture has been proposed as a potential technique for extracting useful information from  complicated datasets pertaining to the SB. Local processors in SBs provide limited computing power to deal with the most sensitive information, such as human-related data. Edge computing, with more computing resources and less latency, can help to satisfy computing tasks with high-reliability and low-latency requirements. Cloud computing equipped with almost infinite computing resources could eventually be played as the `trump card' for particularly computation-hungry tasks. SB-Edge-Cloud computing architecture has much promise for intelligent applications in SBs; however, scheduling the offloading of tasks remains a substantial challenge for researchers.

\subsection*{Realistic Factors Affecting the Implementation}
The above description of the proposed framework demonstrates that high-rate data collection, transmission, and pre-processing/processing over multiple functional layers might result in significant challenges to reliability, stability, and security. These challenges become more severe when collected data are subject to pollution, malicious user behavior, and active network attacks. Consequently, strictly regulated anomaly detection mechanisms must be involved to ensure the reliability, stability, and security of the entire data network. This topic awaits  future research.

A large number of sensors constitute the sensing layer. By implementing such a framework, all residents and their living conditions are observable and can be `seen' by high-layer programs and processing procedures. This issue risks compromising the privacy of residents. To further promote the proposed framework, further investigation and legislation should be conducted and developed with the aim of ensuring a sufficiently secure data protection mechanism.

\section*{Conclusions}
To meet the rapid trend of urbanization, SB plays an invaluable role. With the aim of equipping SBs with environmental perception and logic reasoning abilities, we envisioned a novel SB framework in the context of 6G communications. Two key technologies of 6G communications, i.e., indoor VLC and ML, were jointly applied to construct a reliable transmission infrastructure and perform big data analytics while adapting the indoor environment of the SB. Within this framework, the SB is envisioned to provide a variety of advanced services to residents in a smart and efficient manner. To promote further research and implement the framework in practice, we also simulated a simplistic case to verify its feasibility and considered the challenges facing such SBs and potential future research directions to mitigate these challenges.

\bibliographystyle{IEEEtran}
\bibliography{bib}

\begin{IEEEbiographynophoto}{Shuping Dang} [M'18] (shuping.dang@kaust.edu.sa) received a B.Eng (Hons) in Electrical and Electronic Engineering from the University of Manchester (with first class honors) and a B.Eng in Electrical Engineering and Automation from Beijing Jiaotong University in 2014 via a joint `2+2' dual-degree program, and a D.Phil in Engineering Science from University of Oxford in 2018. He is currently working as a Postdoctoral Fellow with the Computer, Electrical and Mathematical Science and Engineering (CEMSE) Division, King Abdullah University of Science and Technology (KAUST).
\end{IEEEbiographynophoto}
\begin{IEEEbiographynophoto}{Guoqing Ma}[S'18] (guoqing.ma@kaust.edu.sa) received B.Eng in South University of Science and Technology of China (SUSTC) in 2017. He is currently pursuing his Ph.D degree in the M.S/Ph.D program with the Computer, Electrical and Mathematical Science and Engineering (CEMSE) Division, King Abdullah University of Science and Technology (KAUST).
\end{IEEEbiographynophoto}
\begin{IEEEbiographynophoto}{Basem Shihada}[SM'12] (basem.shihada@kaust.edu.sa) is an associate and founding professor of computer science and electrical engineering in the Computer, Electrical and Mathematical Sciences and Engineering (CEMSE) Division at King Abdullah University of Science and Technology (KAUST). Before joining KAUST in 2009, he was a visiting faculty at the Computer Science Department in Stanford University. His current research covers a range of topics in energy and resource allocation in wired and wireless communication networks, including wireless mesh, wireless sensor, multimedia, and optical networks. He is also interested in SDNs, IoT, and cloud computing. In 2012, he was elevated to the rank of Senior Member of IEEE.
\end{IEEEbiographynophoto}
\begin{IEEEbiographynophoto}{Mohamed-Slim Alouini}[F'09] (slim.alouini@kaust.edu.sa) received the Ph.D. degree in electrical engineering from the California Institute of Technology (Caltech), Pasadena, CA, USA, in 1998. He served as a faculty member at the University of Minnesota, Minneapolis, MN, USA, then at Texas A\&M University at Qatar, Education City, Doha, Qatar, before joining King Abdullah University of Science and Technology (KAUST), Thuwal, Makkah Province, Saudi Arabia as a professor of electrical engineering in 2009. At KAUST, he leads the Communication Theory Lab and his current research interests include the modeling, design, and performance analysis of wireless communication systems.
\end{IEEEbiographynophoto}

\end{document}